\documentclass[twocolumn,aps,prl,floats,floatfix]{revtex4}
\usepackage{amsmath}
\usepackage{graphicx}
\usepackage{amssymb}

\usepackage{amsfonts}
\usepackage{graphicx}

\begin{document}

\title{Coherent molecular bound states of bosons and fermions
 near a Feshbach resonance}

\author{P. D. Drummond}

\affiliation{ARC Centre of Excellence for Quantum-Atom Optics, Department of Physics, University of Queensland, Brisbane, Qld 4072, Australia}

\author{K. V. Kheruntsyan}

\affiliation{ARC Centre of Excellence for Quantum-Atom Optics, Department of Physics, University of Queensland, Brisbane, Qld 4072, Australia}

\date{\today{}}

\begin{abstract}
We analyze molecular bound states of atomic quantum gases near a
Feshbach resonance. A simple, renormalizable field theoretic model
is shown to have exact solutions in the two-body sector, whose
binding energy agrees well with observed experimental results in
both Bosonic and Fermionic cases. These solutions, which
interpolate between BEC and BCS theories, also provide a more
general variational ansatz for resonant superfluidity and related
problems.

PACS numbers: 03.75.Nt, 05.30-d, 39.25.+k, 67.40.-w
\end{abstract}
\maketitle The coherent transformation of a cold atomic gas to
molecules in the vicinity of a photoassociation \cite{Heinzen-exp}
or Feshbach \cite{Donley} resonance has enabled a fascinating
probe of quantum dynamical behavior in coupled atom-molecular
systems, together with remarkably precise measurements of quantum
binding energies. Recent Bosonic experiments have extended the
available species to $^{133}$Cs, $^{87}$Rb, and $^{23}$Na
\cite{Bosonic-Cs-Rb-Na}. Experiments on ultracold degenerate Fermi
gases of $^{40}$K and $^{6}$Li atoms have resulted in spectacular
demonstrations of molecular Bose-Einstein condensate (BEC)
formation \cite{Regal,Fermionic-other} and of possible fermion
superfluid behavior in the BEC-BCS crossover region \cite{BCS}.

Since these are many-body systems, it is useful to try to develop
the simplest possible field-theoretic model that can explain their
behavior. An essential feature of any correct many-body treatment
is that the basic theory must be able to reproduce the physics of
the two-body interactions. In this paper, we combine previous
analytic solutions of a coherently coupled field theory
\cite{DKH98,KD98,KD-nondegenerate} with an exact renormalization
of the coupling constants \cite{Holland-Kokkelmans-2001-2002}, in
order to obtain analytic predictions for the two-body bound
states. This gives a unified picture of any Feshbach resonance
experiment and related studies \cite{DKH98,KD98,
KD-nondegenerate,HWDK2000,Timmermans,Javanainen-Mackie-1999-2002,Verhaar,
Yurovski-Julienne,Holland-Kokkelmans-2001-2002,Goral-Rzazewski-2001,
Duine-Stoof-JOptB,Duine-Stoof-review,Koehler-Burnett-2003,Bruun-Pethick},
provided a small number of observable parameters are available.
The predictions will be compared with experimental data and with
coupled-channel calculations.

To quantitatively model these experiments, consider an effective
Hamiltonian for the molecular field ($\hat{\Psi}_{0}$) in the
closed channel and the atomic fields ($\hat{\Psi}_{1(2)}$) in the
free-atom dissociation limit of the open channel:
\begin{equation}
\hat{H}_{1}=\hat{H}_{0}+\frac{\hbar}{{2}}\int d^{3}\mathbf{x}\sum\nolimits _{i,j}U_{ij}\hat{\Psi}_{i}^{\dag}\hat{\Psi}_{j}^{\dag}\hat{\Psi}_{j}\hat{\Psi}_{i},\label{eq:freeham}
\end{equation}
with the commutation (+) or anti-commutation (-) relation
$[\hat{\Psi}_{i}(\mathbf{x},t),\hat{\Psi}_{j}^{\dagger}(\mathbf{x}^{\prime},t)]_{\pm}=\delta_{ij}\delta(\mathbf{x}-\mathbf{x}^{\prime})$
for Bosonic or Fermionic field operators $\hat{\Psi}_{i}$,
respectively. The free Hamiltonian $\hat{H}_{0}$ includes the
usual kinetic energy terms and the potential energies (including
internal energies) due to the trap potential $\hbar V_{i}(x)$,
while $U_{ij}$ is the atom-atom, atom-molecule, and
molecule-molecule coupling due to $s$-wave scattering. The atomic
and molecular masses are $m_{1,2}$ and $m_{0}=m_{1}+m_{2}$, and
$E_{m}=\hbar\left[V_{0}(0)-V_{1}(0)-V_{2}(0)\right]$ gives the
bare energy detuning of the molecular state with respect to free
atoms.

Next, we consider a coherent process of Raman photoassociation or
a magnetic Feshbach resonance coupling, giving rise to an overall
effective Hamiltonian term in the homonuclear case (only with
bosons) \cite{DKH98,KD98}
\begin{equation}
\hat{H}=\hat{H}_{1}+\frac{\hbar\chi}{2}\int d^{3}\mathbf{x}\left[\hat{\Psi}_{0}^{\dagger}\hat{\Psi}_{1}^{2}+\hat{\Psi}_{1}^{\dagger\,2}\hat{\Psi}_{0}\right],\label{eq:homogcouple}\end{equation}
 or, for the case of heteronuclear dimer formation involving either fermions or bosons \cite{KD-nondegenerate}:
\begin{equation}
\hat{H}=\hat{H}_{1}+\hbar\chi\int d^{3}\mathbf{x}\left[\hat{\Psi}_{0}^{\dagger}\hat{\Psi}_{1}\hat{\Psi}_{2}+\hat{\Psi}_{2}^{\dagger}\hat{\Psi}_{1}^{\dagger}\hat{\Psi}_{0}\right].\label{eq:heterocouple}
\end{equation}
Here, $\chi$ is the bare atom-molecule coupling responsible for
the conversion of free atom pairs into molecules and vice versa.
The heteronuclear case can be applied to Fermionic atom pairs in
different spin states ($\hat{\Psi}_{1}$, $\hat{\Psi}_{2}$)
combining into a Bosonic molecule ($\hat{\Psi}_{0}$), or pairs of
Bosonic and Fermionic atoms combining into a Fermionic molecule,
or else to a fully Bosonic case where the atom pairs are not
identical.

\textit{Bosonic homonuclear case}. First we consider the fully
Bosonic uniform case of Eq. (\ref{eq:homogcouple}), i.e., a
single-species atomic BEC (with $m_{1}\equiv m$) coupled to a
molecular BEC, where the atomic background energy is chosen to be
zero. We ignore inelastic collisions -- which is a reasonable
approximation at low density, and let $\kappa=U_{11}$, where
$\kappa$ is the bare atom-atom coupling due to $s$-wave
scattering.

Here a momentum cutoff is implicitly assumed, since in
renormalizable theories one expects to obtain finite results only
after the infinities are absorbed through a redefinition of bare
couplings. To manipulate integrals that \textit{a priori} are
divergent, we regularize them by a simple cutoff: integrals over
$\mathbf{k}$ are restricted to $|\mathbf{k}|<K$.

The homogeneous Hamiltonian, Eq. (\ref{eq:homogcouple}), has an
exact eigenstate in the simplest two-particle sector
\cite{DKH98,KD98}. In momentum space, we expand the field
operators $\hat{\Psi}_{0}(\mathbf{x})$ and
$\hat{\Psi}_{1}(\mathbf{x})$ in terms of Fourier components
$\hat{a}(\mathbf{k)}$ and $\hat{b}(\mathbf{k})$, respectively,
with commutation relations
$[\hat{a}(\mathbf{k}),\hat{a}^{\dagger}(\mathbf{k}^{\prime})]=[\hat{b}(\mathbf{k}),\hat{b}^{\dagger}(\mathbf{k}^{\prime})]=\delta(\mathbf{k}-\mathbf{k}^{\prime})$.
Including a cutoff $K$, the (unnormalized) two-particle eigenstate
corresponding to the zero center-of-mass momentum is given by
\cite{DKH98,KD98}
\begin{equation} \left|\Psi^{(N)}\right\rangle
=\left[\hat{a}^{\dagger}(0)+\int\nolimits
_{|\mathbf{k}|=0}^{K}\frac{d^{3}\mathbf{k\,}G(\mathbf{k})}{{(2\pi)^{3/2}}}\hat{b}^{\dagger}(\mathbf{k})\hat{b}^{\dagger}(-\mathbf{k})\right]^{N/2}\left|0\right\rangle
,\label{Eigenstate}
\end{equation}
where $N=2$ in the exactly soluble two-particle case, and
$G(\mathbf{k})$ is the atomic pair correlation function in Fourier
space. This coherent superposition of a molecule with correlated
pairs of atoms can be viewed as a dressed molecule. More
generally, this is also a useful low-density variational ansatz
for $N>2$ particles, where it describes a BEC of dressed molecules
\cite{DKH98,KD98}.

\textit{Fermionic or heteronuclear case.} Next, we wish to
consider the important case of Fermionic atom pairs (with
$m_{1}=m_{2}\equiv m$) in different spin states combining into a
Bosonic molecule. This is especially relevant to the studies of
ultracold Fermi gases \cite{Regal,Fermionic-other,BCS} in the
region of resonant superfluidity and BEC-BCS crossover. These
experiments are notable for the greatly reduced inelastic loss
rate from atom-molecular collisions, due to Pauli blocking
\cite{Shlyapnikov}. In this Fermionic case, we only have an
$s$-wave coupling between the different fermions, so that
$\kappa=U_{12}$. In addition, the final results of this section
can be applied to heteronuclear molecules (with either statistics
of the constituent atoms), except that the mass $m$ has to be
replaced by $2m_{r},$ where $m_{r}=m_{1}m_{2}/(m_{1}+m_{2})$ is
the reduced mass.

The Hamiltonian (\ref{eq:heterocouple}) relevant to this case,
also has an exact eigenstate in the two-particle ($N=2$) sector
\cite{KD-nondegenerate}. Expanding the field operators
$\hat{\Psi}_{1,2}(\mathbf{x})$ in terms of Fourier components
$\hat{b}_{1,2}(\mathbf{k})$, the eigenstate is now given by
\begin{equation}
\left|\Psi^{(N)}\right\rangle=\left[\hat{a}^{\dagger}(0)+\int\nolimits
_{|\mathbf{k}|=0}^{K}\frac{d^{3}\mathbf{k\,}G(\mathbf{k})}{{(2\pi)^{3/2}}}\hat{b}_{1}^{\dagger}(\mathbf{k})\hat{b}_{2}^{\dagger}(-\mathbf{k})\right]^{N/2}\left|0\right\rangle
.\label{Eigenstate2}
\end{equation}
As before, this is also a useful variational ansatz for the
$N$-particle ($N>2$) ground state, where it extends BCS theory to
include a coherent molecular field.

\textit{Exact eigenvalues. }In either the homonuclear or
heteronuclear case, the exact energy eigenvalue corresponding to
the two-particle eigenstate ($N=2$) is known
\cite{KD98,KD-nondegenerate}. Introducing a multiplicity parameter
$s$, where $s=1$ for the homonuclear case, and $s=2$ for the
Fermionic or heteronuclear case, we find that
\begin{equation}
E=E_{m}-\frac{\hbar s\chi^{2}}{2}\left[\kappa+\frac{2\pi^{2}\hbar
r_{0}/m}{r_{0}K-\tan^{-1}(r_{0}K)}\right]^{-1}=-\frac{\hbar^{2}}{mr_{0}^{2}}.\label{E}
\end{equation}
For real and positive $r_{0}$, this corresponds to a bound state
with negative energy, and the resulting binding energy is
$E_{b}=-E$. The quantity $r_{0}$ is the correlation radius or the
effective size of the dressed molecule. The right-hand side of Eq.
(\ref{E}) needs to be solved for $r_{0}$, or equivalently for the
binding energy $E_{b}\equiv\hbar^{2}/(mr_{0}^{2})$ as a function
of $E_{m}$, but in general it has no explicit solution.

Next, it is useful to re-express the bare Hamiltonian parameters
in terms of renormalized observable parameters that are invariant
at large momentum cutoff. We therefore include a nonperturbative
renormalization using integral equation methods from scattering
theory \cite{Holland-Kokkelmans-2001-2002}, which has some subtle
features. In particular, a repulsive contact potential with
$\kappa>0$ has no effect -- it does not lead to scattering in
three-dimensional field theory. However, either positive or
negative scattering lengths can be generated from the same type of
attractive contact potential with $\kappa<0$, depending on the
limiting procedure: if it is carried out with sufficiently deep
potentials to allow a bound state to form in the atomic field
channel, then a positive scattering length is possible even with
an attractive short-range potential.

The renormalization \cite{Holland-Kokkelmans-2001-2002} expresses
the bare values as $\kappa=\Gamma\kappa_{0}$,
$\chi=\Gamma\chi_{0}$, and
$E_{m}=E_{0}+s\hbar\beta\Gamma\chi_{0}^{2}/2$, in terms of the
observed or renormalized values $\kappa_{0}$, $\chi_{0}$, and
$E_{0}$. In the Feshbach resonance case, for definiteness,
$E_{0}=\Delta\mu(B-B_{0})$. Here, the cutoff $K$ is included
through a scaling parameter
$\Gamma=\left(1-\beta\kappa_{0}\right)^{-1}$, where
$\beta=mK/(2\pi^{2}\hbar)$, $\kappa_{0}=4\pi\hbar a_{bg}/m$, and
$a_{bg}$ is the background $s$-wave scattering length for the
atoms. In the homonuclear case, $\Delta\mu=2\mu_{1}-\mu_{m}$ is
the magnetic moment difference between the atomic and the bound
molecular channels, and $B_{0}$ is the magnetic field
corresponding to the resonance, while in the heteronuclear case,
$\Delta\mu=\mu_{1}+\mu_{2}-\mu_{m}$.

We now wish to rewrite Eq. (\ref{E}) in terms of the renormalized
constants $\chi_{0}$, $\kappa_{0}$, and $E_{0}$. After taking the
limit of large momentum cutoff $K$, we obtain the following simple
analytic result:
\begin{equation}
E_{0}=-E_{b}-\frac{sC\hbar\chi_{0}^{2}\sqrt{E_{b}}}{1-2C\kappa_{0}\sqrt{E_{b}}},\label{E-0}
\end{equation}
where $C\equiv m^{3/2}/(8\pi\hbar^{2})$. Using
$E_{0}=\Delta\mu(B-B_{0})$, Eq.~(\ref{E-0}) can also be rewritten
in terms of the magnetic fields, so that the resulting binding
energy can be directly compared with the experimental data.

In the JILA $^{85}$Rb experiments
\cite{Donley,Claussen-high-precision}, the creation of a
homonuclear dressed molecular state at a given $B$ value is
followed by a rapid change in the magnetic field which allows an
interference fringe to be observed in the remaining total number
of atoms. The reason for the fringe is due to the fact that in a
dynamical experiment, paired atoms in the dressed molecular wave
function can interfere constructively or destructively with
condensate atoms that are not in the ground-state wave function.
The resulting interference pattern oscillates with a frequency
corresponding to the dressed molecular binding energy.

\begin{figure}
\includegraphics[%
  width=7cm]{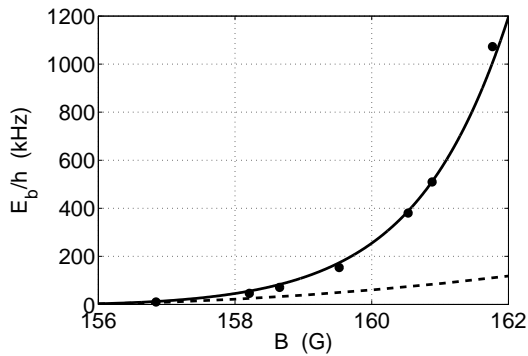}

\caption{Binding energy $E_{b}$ versus the magnetic field $B$ for
the $^{85}$Rb experiment \protect\cite{Donley}. The parameter
values are taken from the subsequent high-precision measurements
on the same system \protect\cite{Claussen-high-precision}:
$B_{0}=155$ G and $a_{bg}=-443a_{0}$, where $a_{0}$ is the Bohr
radius. In addition, we take $\chi_{0}=2.84\times10^{-4}$
m$^{3/2}$/s and $\Delta\mu=-2.23\mu_{B}$
\protect\cite{Holland-private}, where $\mu_{B}$ is the Bohr
magneton. The solid line is our theoretical result, Eq. (\protect
\ref{E-0}), while the dashed line is the result of Eq.
(\protect\ref{E-b-Stoof}). The circles are the experimental data
of Ref. \protect\cite{Claussen-high-precision}.}

\label{fig:Binding-energy-Rb85}
\end{figure}

A graphical solution of Eq. (\ref{E-0}), i.e., the binding energy
$E_{b}$ vs $B$ for the JILA $^{85}$Rb experiment
\cite{Donley,Claussen-high-precision}, is plotted in Fig.
\ref{fig:Binding-energy-Rb85}, together with experimentally
observed Ramsey fringe frequencies, which are interpreted in the
experiment as a dressed molecular binding energy. The agreement
between this simple analytic result and the experimentally
observed binding energy (as well as the coupled-channel
calculation \cite{Holland-Kokkelmans-2001-2002}) is excellent.

\begin{figure}
\includegraphics[%
  width=7cm]{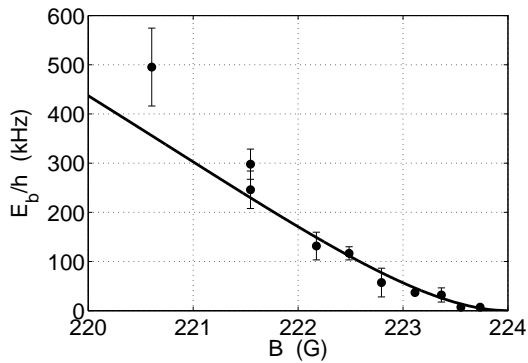}

\caption{Binding energy $E_{b}$ vs $B$ for the $^{40}$K experiment
\protect\cite{Regal}, where $B_{0}=224$ G and $a_{bg}=174a_{0}$.
In addition, we take $\chi_{0}=1.12\times10^{-4}$ m$^{3/2}$/s and
$\Delta\mu=1.27\mu_{B}$ \protect\cite{Holland-private}. The solid
line is our theoretical result, while the circles and the error
bars are the experimental data from Fig. 5 of Ref.
\protect\cite{Regal}.}

\label{fig:Binding-energy-K40}
\end{figure}

Binding energy measurements were also carried out for the case of
Fermionic $^{40}$K atoms in two different spin states combining
into a Bosonic molecule \cite{Regal}. In Fig.
\ref{fig:Binding-energy-K40} we plot the solution to Eq.
(\ref{E-0}), i.e., the binding energy $E_{b}$ vs the magnetic
field $B$, for this experiment, where we also see a good agreement
between the theory and experiment.

\emph{Near-threshold physics.} There are common features with
either Fermionic and Bosonic atoms. All results are expressed in
terms of the four observable parameters $\chi_{0}$, $\kappa_{0}$,
$\Delta\mu$, $B_{0}$, and are clearly independent of the cutoff,
as one would expect from a renormalizable theory. There are two
cases corresponding to different signs of $\kappa_{0}$:

(i) Attractive case. If $a_{bg}<0$, then $E_{b}$ is a
single-valued function of $B$, so there is only one solution
branch.

(ii) Repulsive case. If $a_{bg}>0$, then $E_{b}$ is a
double-valued function of $B$, so there are two solution branches.
This case has a bound state in the atomic channel.

In all cases, the physics near threshold is crucial to
understanding either type of experiment. For small binding
energies in the vicinity of the resonance, Eq. (\ref{E-0}) gives a
quadratic dependence of $E_{b}$ on $E_{0}$ (or on the magnetic
field $B$): $E_{b}\simeq E_{0}{}^{2}/(sC\hbar\chi_{0}^{2})^{2}$,
where $E_{0}=\Delta\mu(B-B_{0})$. This is in agreement with the
simple resonant scattering theory result that
$E_{b}=\hbar^{2}/ma(B)^{2}$ near the resonance
\cite{Sakurai,Duine-Stoof-review}. Here, the effective scattering
length is $a(B)=a_{bg}\left[1-\Delta B/(B-B_{0})\right]$, $\Delta
B$ is the width of the resonance, and the atom-molecule coupling
$\chi_{0}$ can be expressed via $\Delta B$ as follows:
$\chi_{0}\simeq\sqrt{8\pi a_{bg}\Delta\mu\Delta B/(sm)}$.

In the opposite limit of large $E_{b}$, i.e., for magnetic fields
far away from the resonance, the same equation (\ref{E-0}) gives
linear dependence of $E_{b}$ on $E_{0}$ (and hence on $B$) as
expected, $E_{b}\simeq-E_{0}+s\hbar\chi_{0}^{2}/(2\kappa_{0})$,
including a constant energy shift. This linear behavior is not
accessible with the resonant scattering theory result of
$E_{b}=\hbar^{2}/ma(B)^{2}$.

For $C|\kappa_{0}|\sqrt{E_{b}}\ll1$, i.e., either for small
background scattering $|a_{bg}|$ or small binding energies $E_{b}$
near the resonance, we can neglect $2C\kappa_{0}\sqrt{E_{b}}$ in
the denominator of the second term in Eq. (\ref{E-0}) and obtain a
quadratic with respect to $\sqrt{E_{b}}$. This has the following
explicit solution:
\begin{equation}
E_{b}\simeq-E_{0}-\frac{(sC\hbar\chi_{0}^{2})^{2}}{2}\left[\sqrt{1-\frac{4E_{0}}{(sC\hbar\chi_{0}^{2})^{2}}}-1\right],\label{E-b-Stoof}
\end{equation}
which (for $s=1$) coincides with Eq. (21) of Ref.
\cite{Duine-Stoof-JOptB}. This result formally incorporates the
above-mentioned quadratic dependence of $E_{b}$ on $E_{0}$ near
the resonance where $4E_{0}/(sC\hbar\chi_{0}^{2})^{2}\ll1$, and
the linear dependence far away from the resonance. The quadratic
dependence is in qualitative agreement with the behavior found
from our exact result. However, the linear part
($E_{b}\simeq-E_{0}$) -- while giving the correct slope of the
binding energy -- does not account for the energy shift term
$s\hbar\chi_{0}^{2}/(2\kappa_{0})$ due to the renormalization of
$E_{m}$. This leads to a discrepancy seen in Fig.
\ref{fig:Binding-energy-Rb85} (dashed line) away from the
resonance, and is due to the fact that the assumption of
$C|\kappa_{0}|\sqrt{E_{b}}\ll1$ used to obtain Eq.
(\ref{E-b-Stoof}) is inconsistent with the case of large binding
energies under consideration. We note here that it is also
possible to obtain the exact result from the molecular Green's
function method of Ref. \cite{Duine-Stoof-JOptB}, if the relevant
self-energy term is included without approximation \cite{Liu}.

The relative fraction of the atomic and molecular components in
the two-particle eigenstate can be calculated using the conserved
total number of atomic particles,
$\hat{N}=\hat{N}_{1}+2\hat{N}_{0}$ (or
$\hat{N}=\hat{N}_{1}+\hat{N}_{2}+2\hat{N}_{0}$ in the Fermionic
case). At low density, the closed-channel molecular fraction is
(including a factor of 2 to reflect the fact that each molecule
consists of two atoms): \begin{equation}
2N_{0}/N=\left(1+2F/s\right)^{-1}.\label{molecule-fraction}
\end{equation}
Here, $F\equiv\int d^{3}\mathbf{k}G^{2}(\mathbf{k})/(2\pi)^{3}$,
and the correlation function $G(\mathbf{k})$ is given by a
Lorentzian
\begin{equation}
G(\mathbf{k})=G_{0}/\left(1+r_{0}^{2}k^{2}\right),\label{Gk}
\end{equation}
where $k=|\mathbf{k}|$,
$G_{0}=sm\chi_{0}r_{0}^{3}/[2\hbar(a_{bg}-r_{0})]$, and we have
already taken into account renormalization. Taking the integral in
$F$ we obtain that
\begin{equation} F=\frac{G_{0}^{2}}{8\pi
r_{0}^{3}}=\frac{s^{2}m^{3/2}\chi_{0}^{2}}{32\pi\hbar\sqrt{E_{b}}}\left(1-\frac{m^{3/2}\kappa_{0}}{4\pi\hbar^{2}}\sqrt{E_{b}}\right)^{-2},\label{F-Eb}
\end{equation}
where we have expressed $r_{0}=\hbar/\sqrt{mE_{b}}$, so that the
final result can be analyzed as a function of the magnetic field
$B$ using the solution to Eq. (\ref{E-0}).

Combining this result with Eqs. (\ref{E-0}) and
(\ref{molecule-fraction}), we find that the average fraction of
bare molecules in the closed channel is typically very small near
the resonance. For example, for $^{85}$Rb parameters
\cite{Claussen-high-precision} it is no higher than
$2N_{0}/N\simeq0.07$, for magnetic fields from $B_{0}$ to
$B\simeq160$ G. This implies that the structure of the dressed
molecules and the underlying physics near the resonance is
dominated by the correlated atom pairs rather than by the closed
channel molecules.

We can also calculate the atomic pair correlation in coordinate
space. This is the inverse Fourier transform of $G(\mathbf{k})$
given by $g(\mathbf{x})=G_{0}\exp(-|\mathbf{x}|/r_{0})/(2\pi
r_{0}^{2}|\mathbf{x}|)$, for $|\mathbf{x|}>0$. Since
$r_{0}=\hbar/\sqrt{mE_{b}}$, it is clear that, near threshold, the
bound states are superpositions of molecules with pairs of atoms
at very long range. Here, one can expect modifications
\cite{Duine-Stoof-review} of the binding energy due to mean-field
many-body interactions of the correlated atoms, which have a
character similar to Cooper pairs. Such departures from the
predicted binding energies are indeed observed
\cite{Claussen-high-precision} in high-precision Ramsey
spectroscopy for $^{85}$Rb. Similarly, recent collective-mode
spectroscopy in $^{6}$Li has revealed BCS-like behavior near
threshold with reduced mode frequencies \cite{Grimm}, quite
different to that expected for a conventional molecular BEC.

In summary, a relatively simple field-theoretic model for Feshbach
coupling has exact solutions for the eigenstates in the
low-density two-particle sector. It is able to accurately predict
Feshbach dressed-molecule binding energies, and also gives a
physical understanding of the type of correlated atom-molecular
structure produced in these experiments. The model has a simple,
universal character, and can be used to describe a variety of
cases with both positive and negative background scattering
length. Having analytic solutions of the actual eigenstates
provides an alternative picture that aids in understanding these
interesting experiments, and is readily usable as a starting point
to a more complete many-body theory.

The authors gratefully acknowledge the Australian Research Council
for the support of this work, and thank M. Holland, C. Pethick,
and X-J. Liu for useful discussions. The research was also
supported by the National Science Foundation under Grant No.
PHY99-07949.

\end{document}